\documentclass[a4paper,12pt]{jpconf}
\usepackage[latin9]{inputenc}
\usepackage{units}
\usepackage{textcomp}
\usepackage{amstext}
\usepackage{graphicx}
\usepackage{esint}
\usepackage[unicode=true,pdfusetitle,
 bookmarks=true,bookmarksnumbered=false,bookmarksopen=false,
 breaklinks=false,pdfborder={0 0 1},backref=false,colorlinks=false]
 {hyperref}

\makeatletter

\pdfpageheight\paperheight
\pdfpagewidth\paperwidth

\usepackage[square,sort&compress,numbers]{natbib}
\usepackage{sidecap}
\sidecaptionvpos{figure}{t}

\usepackage[font=small,format=plain,labelfont=bf,up]{caption}

\makeatother

\begin{document}
\title{Measurements of entanglement over a kilometric distance to test superluminal models of Quantum Mechanics: preliminary results.}

\author{B Cocciaro, S Faetti and L Fronzoni}

\address{Department of Physics Enrico Fermi, Largo Pontecorvo 3, I-56127 Pisa, Italy}

\ead{ b.cocciaro@comeg.it, sandro.faetti@unipi.it, leone.fronzoni@unipi.it}
\begin{abstract}
As shown in the \emph{EPR} paper (Einstein, Podolsky e Rosen, 1935),
Quantum Mechanics is a non-local Theory. The Bell theorem and the
successive experiments ruled out the possibility of explaining quantum
correlations using only local hidden variables models. Some authors
suggested that quantum correlations could be due to superluminal communications
that propagate isotropically with velocity \emph{$v_{t}>c$} in a
preferred reference frame. For finite values of \emph{$v_{t}$} and
in some special cases, Quantum Mechanics and superluminal models lead
to different predictions. So far, no deviations from the predictions
of Quantum Mechanics have been detected and only lower bounds for
the superluminal velocities \emph{$v_{t}$} have been established.
Here we describe a new experiment that increases the maximum detectable
superluminal velocities and we give some preliminary results.
\end{abstract}

\section{\label{sec:Introduction}Introduction}

\begin{SCfigure}[50][h]
 \centering
 \includegraphics[width=0.5\textwidth]{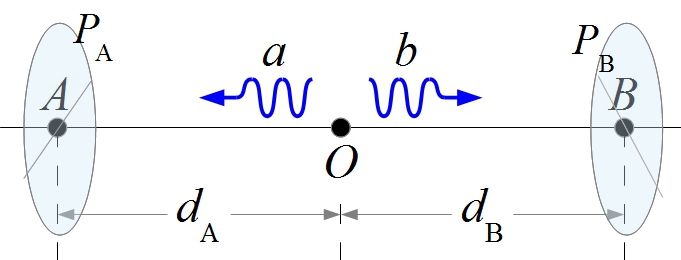}
 \hspace{0.05in}
 \caption{\protect\rule{0ex}{7ex}\textit{O}: source of entangled photons (\emph{a} and \emph{b}); \textit{A}(Alice) and \textit{B}(Bob): points equidistant from \textit{O} ($d_{A}=d_{B}$); \textit{$P_{A}$} and \textit{$P_{B}$}: Polarizing filters centered at points $A$ and $B$, respectively.}
 \label{fig:fotoni entangled}
\end{SCfigure}

The non local character of Quantum Mechanics (\textit{QM}) has been
object of a great debate starting from the famous Einstein-Podolsky-Rosen
(\textit{EPR}) paper \cite{EPR}. Consider, for instance, a quantum
system made by two photons \emph{a} and \emph{b} that are in the polarization
entangled state 

\begin{equation}
|\psi>=\frac{1}{\sqrt{2}}\left(|H,H>+e^{i\phi}|V,V>\right)\label{eq:1}
\end{equation}
where \textit{H} and \textit{V} stand for horizontal and vertical
polarization, respectively, and $\phi$ is a constant phase coefficient.
The two entangled photons are created at point \textit{O}, propagate
in space far away one from the other (see Fig.\ref{fig:fotoni entangled})
and reach at the same time points \textit{A} (Alice) and \textit{B}
(Bob) that are equidistant from \textit{O} as schematically drawn
in Fig.\ref{fig:fotoni entangled}. Two polarizing filters $P_{A}$
and $P_{B}$ lie at points \textit{A} and \textit{B}, respectively.

 Suppose, now, that the polarizers axes are aligned along the horizontal
direction. According to \textit{QM}, the passage of photon \emph{a}
(or \emph{b}) through polarizer $P_{A}$ (or $P_{B}$) leads to the
collapse of the entangled state to $|H,H>$ everywhere, then, also
photon \emph{b} (or \emph{a}) collapses to the horizontal polarization.
This behaviour\textit{ }suggests the existence of a sort of ``action
at a distance'' between entangled particles in complete disagreement
with any other classical physic phenomenon (Electromagnetism, Gravity
....). According to Gisin~\cite{Gisin2014,Gisin_QuantumChance},
classical correlations between far events have always due to two possible
mechanisms: Common Cause or Communications. The Bell theorem~\cite{Bell}
and many successive \emph{EPR} experiments~\cite{Feedman_PhysRevLett_1972,Aspect,Zeilinger_PLA_1986,Tittel_PhysRevLett_1998,Weihs_PhysRevLett_1998,Aspect_Nature_1999,Pan_Nature_2000,Grangier_Nature_2001,Rowe_Nature_2001,Matsukevich_PRL_2008}
demonstrated that correlations cannot be due only to a common cause
(\emph{hidden variables theories}) or to common cause + subluminal
communications. According to Bell, ``\emph{in these EPR experiments
there is the suggestion that behind the scenes something is going
faster than light}''~\cite{Davies_ghost_1993}. Models of \textit{QM}
based on the presence of superluminal communications (tachyons) have
been proposed~\cite{Eberhard_1989,Bohm_undivided_1991}. Tachyons
are known to lead to causal paradoxes (see, for instance, pages 52-53
in~\cite{moller_theory_1955}), but no causal paradox arises if tachyons
propagate isotropically in a preferred frame (\emph{PF}) with velocity
$v_{t}=\beta_{t}c\,(\beta_{t}>1)$~\cite{Kowalczynski_IntJThPhys_1984,Reuse_AnPhys_1984,Caban_PhysRevA_1999,maudlin_quantum_2001,Cocciaro_2013_ShutYourselfUp}. 

Suppose, now, that quantum correlations are due to superluminal communications
and that an \emph{ideal experiment }is performed in the tachyon preferred
frame $S'$ where two polarizing filters lie at the same optical distances
$d'_{A}=d'_{B}$ from source \emph{O}. Photons \emph{a} and \emph{b}
get the polarizers at the same time and no communication is possible.
Then, correlations between entangled particles should differ from
the predictions of \emph{QM} and should satisfy the Bell inequality.
However, from the experimental point of view, equality $d'_{A}=d'_{B}$
can be only approximatively verified within a given uncertainty $\Delta d'$.
Consequently, photons \emph{a} and \emph{b} could get the polarisers
at two different times ($\Delta t'=\nicefrac{\Delta d'}{c}$) and
could communicate if the tachyon velocity exceeds a lower bound $v_{t,min}=c\left(d'_{AB}/\Delta d'\right)$
where $d'_{AB}$ is the distance between polarizers $P_{A}$ and $P_{B}$
in the \emph{PF}. Two are the possible experimental results:\emph{
i)} a lack of quantum correlations is observed; \emph{ii)} quantum
correlations are always satisfied. In the first case (\emph{i)}) one
should conclude that quantum correlations are due to exchange of superluminal
messages with velocity lower than $v_{t,min}$. In the second case
(\emph{ii)}), due to the experimental uncertainty $\Delta d'$, one
cannot invalidate the superluminal model of \emph{QM} but can only
establish a lower bound $v_{t,min}=c\left(d'_{AB}/\Delta d'\right)$
for the superluminal velocities. It has been recently demonstrated
an important theorem~\cite{Bancal_NatPhys_2012,Barnea_PhysRevA2013}:
if \emph{QM} correlations are due to superluminal signals with finite
velocity $v_{t}$, then also a \emph{macroscopic superluminal signalling}
becomes possible provided that states of three or four entangled particles
are involved. This means that the superluminal signals do not remain
hidden but they could lead to macroscopic superluminal communications.
In conclusion, there are two possible alternative situations both
involving some upheaval of the common thought: \emph{a}) Nature is
intrinsically non local and far events can be correlated without any
common cause or communication (orthodox \emph{QM}); \emph{b}) Nature
is local but, in this case, macroscopic superluminal signalling is
possible (superluminal models). Physics is an experimental Science
and, thus, we think that only the experiments can decide between these
two alternatives.

The correlations between entangled particles can be experimentally
tested measuring the number of coincidences $N(\alpha{}_{A},\alpha_{B})$
of photons passing through polarizers $P_{A}$ and $P_{B}$ for different
values of the polarizers angles $\alpha{}_{A}$ and $\alpha_{B}$
with respect to the horizontal axis. In particular, two correlations
parameters $S_{max}$ and $S_{min}$ can be measured (see equations
(33) and (34) in reference~\cite{Aspect_2002}):

\begin{equation}
S_{max}=\frac{N(45\text{\textdegree,67.5\textdegree)- \emph{N}(0\textdegree,67.5\textdegree) -\emph{ N}(45\textdegree,112,5\textdegree) - \emph{N}(90\textdegree,22.5\textdegree)}}{N}\label{eq:Smax}
\end{equation}
and

\begin{equation}
S_{min}=\frac{N(135\text{\textdegree,202.5\textdegree)-\emph{ N}(0\textdegree,202.5\textdegree) -\emph{N}(135\textdegree,157,5\textdegree) - \emph{N}(90\textdegree,67.5\textdegree)}}{N},\label{eq:Smin}
\end{equation}
 where $N$ is the number of coincidences with no polarizers (tacking
into account for the polarizers transmission) that can be written
as:

\begin{equation}
N=N(0\text{\textdegree,0\textdegree)+\emph{N}(0\textdegree,90\textdegree)+\emph{N}(90\textdegree,0\textdegree)+\emph{N}(90\textdegree,90\textdegree).}\label{eq:NTOT}
\end{equation}
Eqs. (\ref{eq:Smax}), (\ref{eq:Smin}) and \ref{eq:NTOT} have been
obtained from equation (33) in reference~\cite{Aspect_2002} using
the equalities:

\begin{equation}
\begin{array}{l}
N(a',\infty)=N(a',b)+N\left(a',b+90\text{\textdegree}\right)\\
N(\infty,b)=N(a,b)+N\left(a+90\text{\textdegree},b\right).
\end{array}\label{eq:Smax-1}
\end{equation}
Quantum Mechanics predicts $S_{max}$ = 0.207 and $S_{min}$ = - 1.207,
respectively, whilst local theories must satisfy the inequalities
$S_{max}$$\leq$ 0 and $S_{min}$ $\geq$ -1. Then, the measurement
of one of these parameters makes possible a direct test of the superluminal
models. 

\begin{SCfigure}[50]
 \centering
 \includegraphics[width=0.5\textwidth]{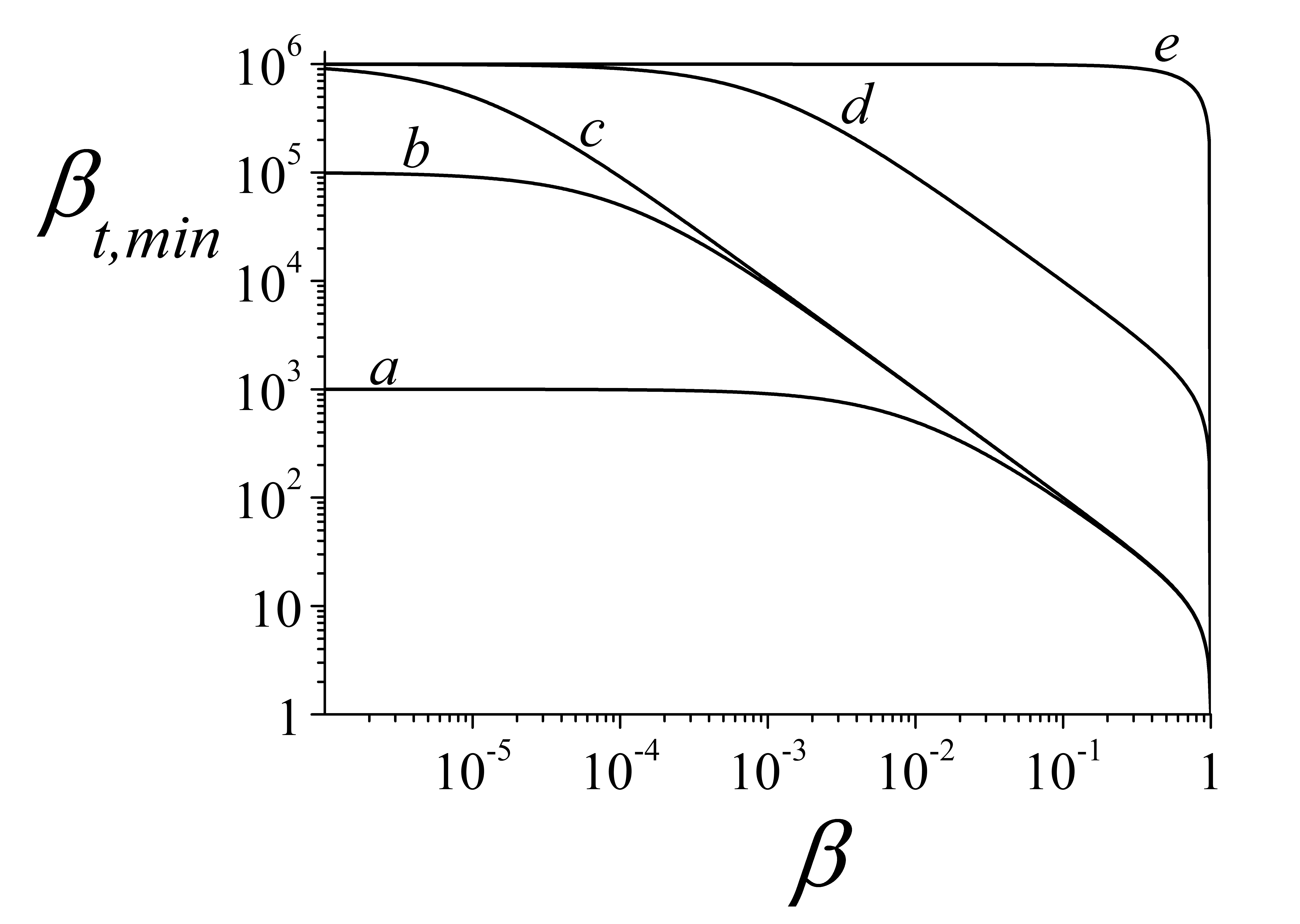}
 \hspace{0.05in}
 \caption{\protect\rule{0ex}{7ex}Function $\beta_{t,min}$ versus $\beta$  for the unfavourable case $\chi=\nicefrac{\pi}{2}$ and for some values of the experimental parameters $\bar{\rho}$ and $\delta t$. Curves \emph{a}, \emph{b} and \emph{c} correspond to the fixed acquisition time $\delta t=10^{-1}\left(\nicefrac{T}{\pi}\right)$ and to decreasing values of $\bar{\rho}$ ($a:\bar{\rho}=10^{-3},\, b:\bar{\rho}=10^{-5},\, c:\bar{\rho}=10^{-6}$). Curves \emph{c}, \emph{d} and \emph{e} correspond to a fixed value $\bar{\rho}=10^{-6}$ and to decreasing values of $\delta t$ ($c:\delta t=10^{-1}\left(\nicefrac{T}{\pi}\right),\, d:\delta t=10^{-3}\left(\nicefrac{T}{\pi}\right),\, e:\delta t=10^{-7}\left(\nicefrac{T}{\pi}\right)$).}
 \label{fig:2}
\end{SCfigure}

So far we considered an ideal experiment performed in the preferred
frame, but the\emph{ PF} is unknown. A more complex \emph{EPR} experiment
can be still performed in the Earth if \emph{A} and \emph{B} are aligned
along the Est-West axis and are equidistant (in the Earth frame) from
the photons source at \emph{O}. Of course, the entangled photons get
simultaneously polarizers $P_{A}$ and $P_{B}$ in the Earth reference
frame but not in the \emph{PF}. However, according to Relativity,
these events become simultaneous also in the \emph{PF} if the velocity
vector $\vec{V}=\vec{\beta}c$ of the \emph{PF} is orthogonal to the
\emph{A}-\emph{B} axis (see, for instance, the appendix in~\cite{Cocciaro_DICE2013}).
If the \emph{A}-\emph{B} axis coincides with the East-West direction,
due to the Earth rotation around its axis, there are always two times
$t_{1}$ and $t_{2}$ during each sidereal day where vector $\vec{V}$
becomes orthogonal to the\emph{ A}-\emph{B} axis. If the \emph{A}-\emph{B
}axis makes an angle $\gamma\neq0$ with the Est-West axis, vector
$\vec{V}$ becomes orthogonal to the \emph{A}-\emph{B} axis only if
angle $\theta$ between vector $\vec{V}$ and the Earth polar axis
lies in the interval {[}$\gamma,\pi-\gamma$ {]}. If this condition
is satisfied, a loss of Quantum correlations should be observed at
two given unknown times $t_{1}$ and $t_{2}$ each day if the tachyon
velocity $v_{t}$ is lower than the maximum detectable velocity $v_{t,min}$.
However, there is an other important feature that can reduce the maximum
detectable tachyons velocities in the Earth experiment. In fact, tachyons
get simultaneously the polarizers also in the \emph{PF} only at the
two well defined times $t_{1}$ and $t_{2}$ but the measure of the
coincidences numbers $N(\alpha{}_{A},\alpha_{B})$ is not instantaneous
and requires a finite acquisition time $\delta t$. This produces
a further uncertainty on the equalization of the optical paths that
is an increasing function of the acquisition time and the reduced
velocity $\beta$ of the\emph{ PF}. Using the Relativity theory, it
has been shown~\cite{Salart_nature_2008,Cocciaro_PLA_2011} that
the lower limit of the detectable tachyon velocities in a Earth experiment
is:

\begin{equation}
\beta_{t,min}=\sqrt{1+\frac{\left(1-\beta^{2}\right)\left[1-\bar{\rho}^{2}\right]}{\left[\bar{\rho}+\beta\sin\chi\sin\frac{\pi\delta t}{T}\right]^{2}}},\label{eq:betamin}
\end{equation}
where $\bar{\rho}=\nicefrac{\Delta d}{d_{AB}}$, $\Delta d$ is the
uncertainty on the equalization of the optical paths in the Earth
frame, \emph{T} is the duration of the sidereal day, $\delta t$ is
the acquisition time, $\chi$ is the polar angle between the North-South
axis and velocity \textbf{\emph{$\vec{V}$}} of the \emph{PF} and
$\beta$ is the reduced\emph{ PF} velocity ($\beta=\nicefrac{V}{c}$).
In typical experimental conditions~\cite{Salart_nature_2008,Cocciaro_PLA_2011,Cinesi_PhysRevLett2013},
the acquisition time $\delta t$ is much smaller than the sidereal
day \emph{T }and $\beta_{t,min}$ is a decreasing function of both
$\bar{\rho}$ and $\delta t$ that reaches a minimum value if $\chi=\nicefrac{\pi}{2}$.
$\beta_{t,min}$ is also a decreasing function of $\beta$ that assumes
its maximum value $\beta_{t,min}=\nicefrac{1}{\bar{\rho}}$ for $\beta=0$
and approaches the minimum value $\beta_{t,min}=1$ for $\beta\rightarrow1$.
Our following considerations and figures will be restricted to $\delta t\ll T$
and to the most unfavourable condition $\chi=\nicefrac{\pi}{2}$.
The typical plot of function $\beta_{t,min}$ versus the reduced velocity
$\beta$ of the \emph{PF} for $\chi=\nicefrac{\pi}{2}$ and for some
values of $\bar{\rho}$ and $\delta t$ is drawn in Fig.\ref{fig:2}.

Experiments of this kind have been performed by some groups in the
last years~\cite{Salart_nature_2008,Cocciaro_PLA_2011,Cinesi_PhysRevLett2013}.
In all these experiments no loss of \emph{QM} correlations has been
observed and, thus, only lower bounds $\beta_{t,min}$ for the tachyons
reduced velocities have been established. Recently~\cite{Cocciaro_DICE2013}
we proposed a new experiment to increase the maximum detectable tachyons
velocities by about two orders of magnitude. Here we describe our
improved experimental apparatus and we report some very preliminary
experimental results. The main features of the experiment are discussed
in Section \ref{sec:The-main-features}. The preliminary experimental
results are in Section \ref{sec:Critical-points-and}, whilst the
conclusions are in Section \ref{sec:Conclusions}.

\section{\label{sec:The-main-features}The experimental apparatus and the
main sources of error.}

\subsection{\label{Entangled source}Production and detection of entangled photons.}

The main goal of our experiment is to make parameters $\bar{\rho}=\nicefrac{\Delta d}{d_{AB}}$
and $\delta t$ as smaller as possible to increase the lower bound
$\beta_{t,min}$. Small values of $\bar{\rho}=\nicefrac{\Delta d}{d_{AB}}$
($\bar{\rho}\approx1.8\times10^{-7}$) are obtained using a large
distance $d_{AB}$ ($d_{AB}=1200\thinspace\mathrm{m}$) and a small
uncertainty $\Delta d$ ($\Delta d<220\thinspace\mathrm{\mathrm{\mu}m}$).
A high intensity source of entangled photons provides a high coincidences
rate ($15000\thinspace\mathrm{coinc/s}$) and, thus, a small minimum
acquisition time that is estimated to be $\delta t\approx0.1\thinspace\mathrm{s}$.
The experiment is performed in the ``East-West'' gallery of the
European Gravitational Observatory (\emph{EGO}~\cite{EGO}) of Cascina
that hosts the \emph{VIRGO} experiment on the detection of gravitational
waves. Unfortunately, this gallery makes an angle $\gamma=18\text{\textdegree}$
with the actual East-West axis. Then, vector $\vec{V}$ becomes orthogonal
to the gallery axis at two times $t_{1}$ and $t_{2}$ only if angle
$\theta$ between vector $\vec{V}$ and the Earth polar axis lies
in the interval {[}$\gamma,\pi-\gamma${]}. This means that we do
not look at the entire celestial sphere but only at a $\approx95\%$
fraction of it. In fact, the excluded solid angle is $\Omega=2\int_{_{0}}^{\gamma}2\pi\sin\theta d\theta\approx5\%$
of the $4\pi$ total solid angle. 

\begin{figure}
\centering{}\includegraphics[scale=0.6]{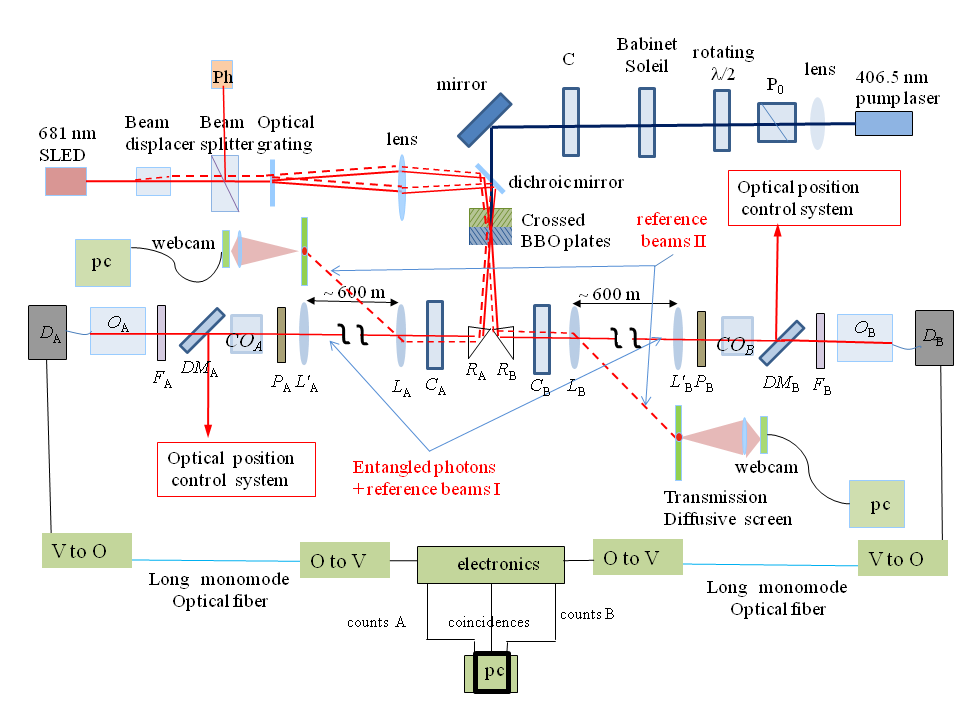}\caption{\label{fig:3} Schematic view of the experimental apparatus. To simplify
the drawing some details that will be discussed below have not be
inserted in the figure. The pump beam (blue in the figure) is polarized
by the polarizer $P_{0}$ and the $\lambda/2$ plate. The Babinet-Soleil
compensator introduces a variable optical dephasing between the horizontal
and vertical polarizations. $C$, $C_{A}$ and $C_{B}$ are anisotropic
compensator plates used to increase the fidelity of the entangled
state. The pump beam is focused on two adjacent \emph{BBO} plates
where the entangled photons are generated and emitted at the angles
$\pm2.42\text{\textdegree}$ with respect to the pump laser beam.
$L_{A},\thinspace L_{B},\thinspace L'_{A}$ and $L'_{B}$ are achromatic
lenses aligned along the \emph{EGO} gallery and having a $6\thinspace\text{m}$
focal length and a $15\thinspace\text{cm}$ diameter. Note that the
figure is not to scale and, in particular, the distances between lenses
$L_{A}\thinspace\mathrm{and}\thinspace L_{B}$ and $\thinspace L'_{A}$
and $L'_{B}$ are very large ($\approx600\thinspace\mathrm{m}$).
$P_{A}$ and $P_{B}$ are polarizing filters. $O_{A}$, $O_{B}$,
$CO_{A}$ and $CO_{B}$ are systems of lenses. $DM_{A}$ and $DM_{B}$
are dichroic mirrors, $F_{A}$ and $F_{B}$ are optical filters, $D_{A}$
and $D_{B}$ are photon counting detectors. The superluminous diode
(\emph{SLED}), the beam displacer and the optical grating are used
to produce two reference beams in each arm of the \emph{EGO} gallery
as discussed below (full and broken red lines). \emph{V} to \emph{O}
denote electronic systems that transform the output voltage pulses
produced by the photon counting detectors into optical pulses, whilst
\emph{O} to \emph{V} transform the optical pulses into voltage pulses. }
\end{figure}
\begin{SCfigure}[50]
 \centering
 \includegraphics[width=0.5\textwidth]{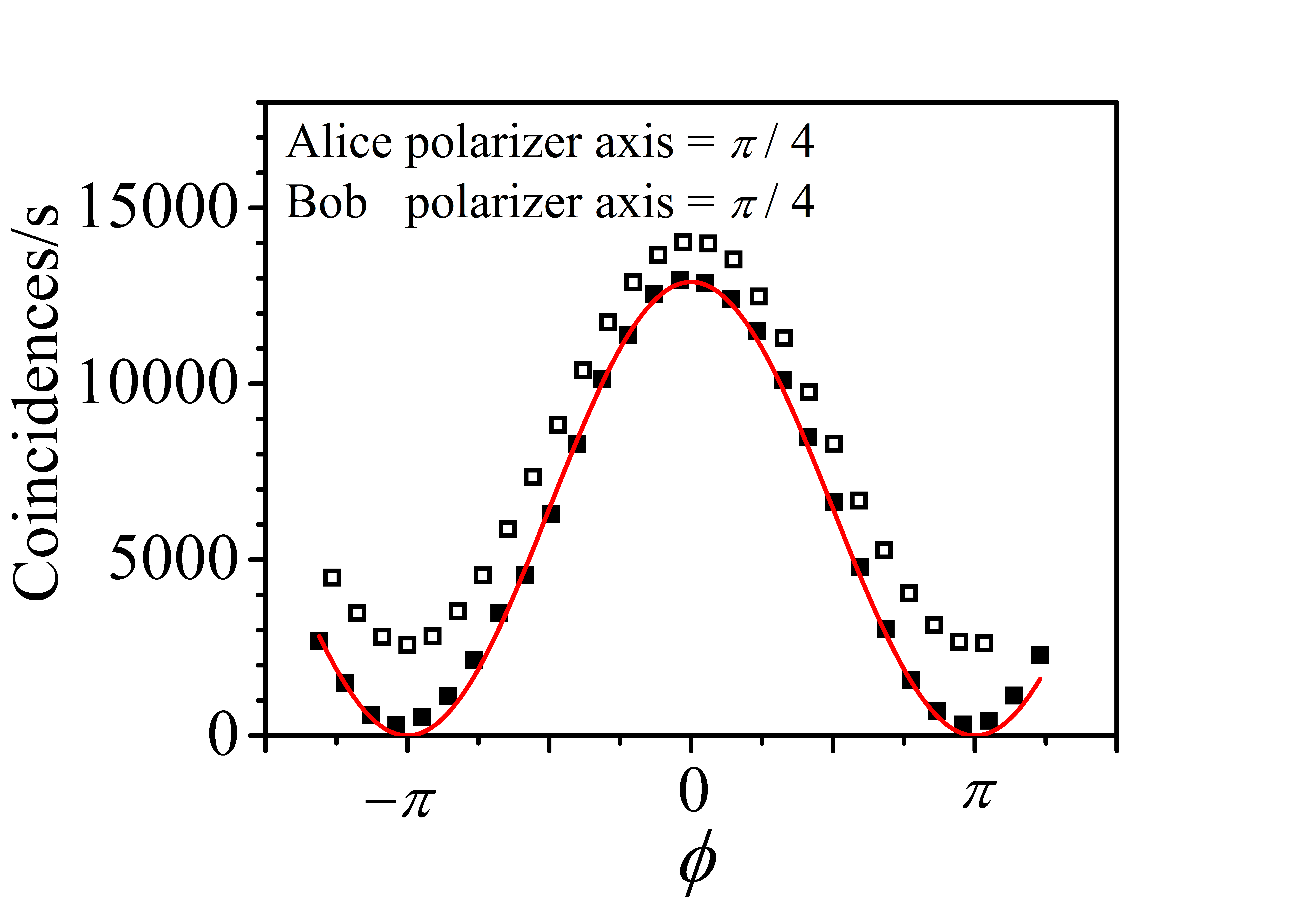}
 \hspace{0.05 in}
 \caption{\protect\rule{0ex}{7ex}Dependence of the coincidences rate on phase $\phi$ when all the polarizers angles are fixed at 45 degrees. Full squares correspond to the experimental values with subctracted statistical coincidences. The open squares are the experimental results with subctracted statistical coincidences but whithout using the Kwiat compensators  \textit{$C_{A}, C_{B} $} and \textit{$C $}. The full line correspond to the prediction of Quantum mechanics for the entangled state of eq.(1).  }
\label{fig:4}
\end{SCfigure}

The experimental apparatus is schematically shown in figure  \ref{fig:3}.
A $220\thinspace\mathrm{mW}$ diode laser beam ($\lambda=406.5\,\mathrm{nm}$)
is polarized (polariser $P_{0}$) and the polarization axis can be
rotated by a motorized $\lambda/2$ plate. All the measurements reported
in this paper have been performed with the polarization axis making
a 45\textdegree{} angle with the horizontal axis. The beam passes
through a Babinet-Soleil compensator and impinges at normal incidence
on two thin ($0.56\,\mathrm{mm}$) adjacent non-linear optical crystals
(\textit{BBO}) cut for type-I phase matching~\cite{Kwiat_PhysRevA_1999}.
The beam is focused on the \emph{BBO }plates with the beam waist having
a $0.6\thinspace\mathrm{mm}$ diameter. The optic axes of the \emph{BBO}
plates are tilted at the angle $29.05\text{\textdegree}$ and lie
in planes perpendicular to each other with the first plane that is
horizontal. The pump beam induces down conversion at the wavelength
$\lambda=813\,\mathrm{nm}$ in each crystal~\cite{Kwiat_PhysRevA_1999}
with maximum emission at the two symmetric angles $\gamma_{A}=-\gamma_{B}=2.42\text{\textdegree}$
with respect to the pump laser beam. Suitable optical diaphragms select
the entangled beams that are emitted within cones of aperture 0.8\textdegree{}
centred at the maximum emission angles. The down converted photons
are created in the maximally entangled state $\left(|H,H>+e^{i\phi}|V,V>\right)/\sqrt{2}$,
where phase $\phi$ can be changed moving the motorized Babinet-Soleil
compensator. Plates $C$, $C_{A}$ and $C_{B}$ in figure \ref{fig:3}
are suitable compensating plates that provide a compensation of spurious
effects due to the poor coherence of the pump beam ($C$) and to the
anisotropy of the \emph{BBO} plates($C_{A}$ and $C_{B}$)~\cite{Kwiat_OptExpr_2005,Kwiat_OptExpr_2007,Kwiat_OptExpr_2009}.
With these compensating plates we obtain a high intensity source of
entangled photons with high fidelity. All the components described
above lie on a central optical table that is entirely enclosed in
an insulating box. One of the lateral internal walls is made by a
$50\thinspace\mathrm{cm\times150\thinspace\mathrm{cm}}$ aluminium
plate ($5\thinspace\mathrm{mm-\mathrm{thickness}}$) in thermal contact
with copper tube coils where a paraflu fluid circulates. Two $80\thinspace\mathrm{W}$
fans inside the box move the air and homogenize the temperature everywhere.
In this way, the internal temperature can be maintained fixed better
than $\pm0.1\thinspace\text{\textdegree C}$. Two couples of specially
designed achromatic lenses $L_{A},\thinspace L_{B},\thinspace L'_{A}$
and $L'_{B}$ (diameter = $15\thinspace\mathrm{cm}$, focal length
= $6\thinspace\mathrm{m}$) allow us to obtain two 1:1 images of the
entangled photons source on two thin near infrared polarizing films
(\emph{LPNIR}, Thorlabs) $P_{A}$ and $P_{B}$ that lie at a $600\thinspace\mathrm{m}$
distance from the source. The entangled photons pass through polarizers
$P_{A}$ and $P_{B}$ and through the optical sets $CO_{A}$ and $CO_{B}$
that will be described below. Then, they are transmitted (98\% transmission)
by dichroic mirrors $DM_{A}$ and $DM_{B}$ (Chroma T760lpxr) and
by two Chroma Techn. Corp. filtering sets $F_{A}$ and $F_{B}$ each
composed by a bandpass filters ET810/40m ($\lambda=810\,\mathrm{nm}\pm20\,\mathrm{nm}$)
and two low pass ET765lp filters ($\lambda_{c}=765\,\mathrm{nm}$).
Finally, two identical optical lenses $O_{A}$ and $O_{B}$ focus
the entangled photons on two Thorlabs multi mode optical fibres having
a large diameter core ($200\,\mathrm{\mu m}$) and high numerical
aperture (0.39). The ends of fibres are connected to the inputs of
the single photons counters $D_{A}$ and $D_{B}$ (Perkin Elmer SPCM-AQ4C)
that generate output voltage pulses with a $25\thinspace\mathrm{ns}$
width. The voltage pulses are transformed into optical pulses by LCM155EW4932-64
modules of Nortel Networks (\emph{V} to \emph{O} module in fig. \ref{fig:3})
that propagate in single mode optical fibres up to the central optical
table where they are converted into electric pulses (\emph{O} to \emph{V}
module in fig.\ref{fig:3}) and sent to an electronic monostable circuit
that provides output squared voltage pulses together with coincidences
pulses. Before starting the measurements we have measured the light
spectral absorption due to air and we have verified that the adsorption
in the wavelengths interval {[}$790\thinspace\mathrm{nm-}830\thinspace\mathrm{nm}${]}
is essentially due to water vapour. The total adsorbed light in this
interval is a fraction lower than 3\% of the incident light for a
45\% air relative humidity. The coincidence rate measured by counters
versus phase $\phi$ is shown in figure \ref{fig:4}. Note the satisfactory
contrast of the fringes that is obtained using the Kwiat et al. compensating
plates~\cite{Kwiat_OptExpr_2005,Kwiat_OptExpr_2007,Kwiat_OptExpr_2009}.

\subsection{\label{reference beams}Compensation of the beam deflections.}

\begin{figure}
\centering{}\includegraphics[scale=0.06]{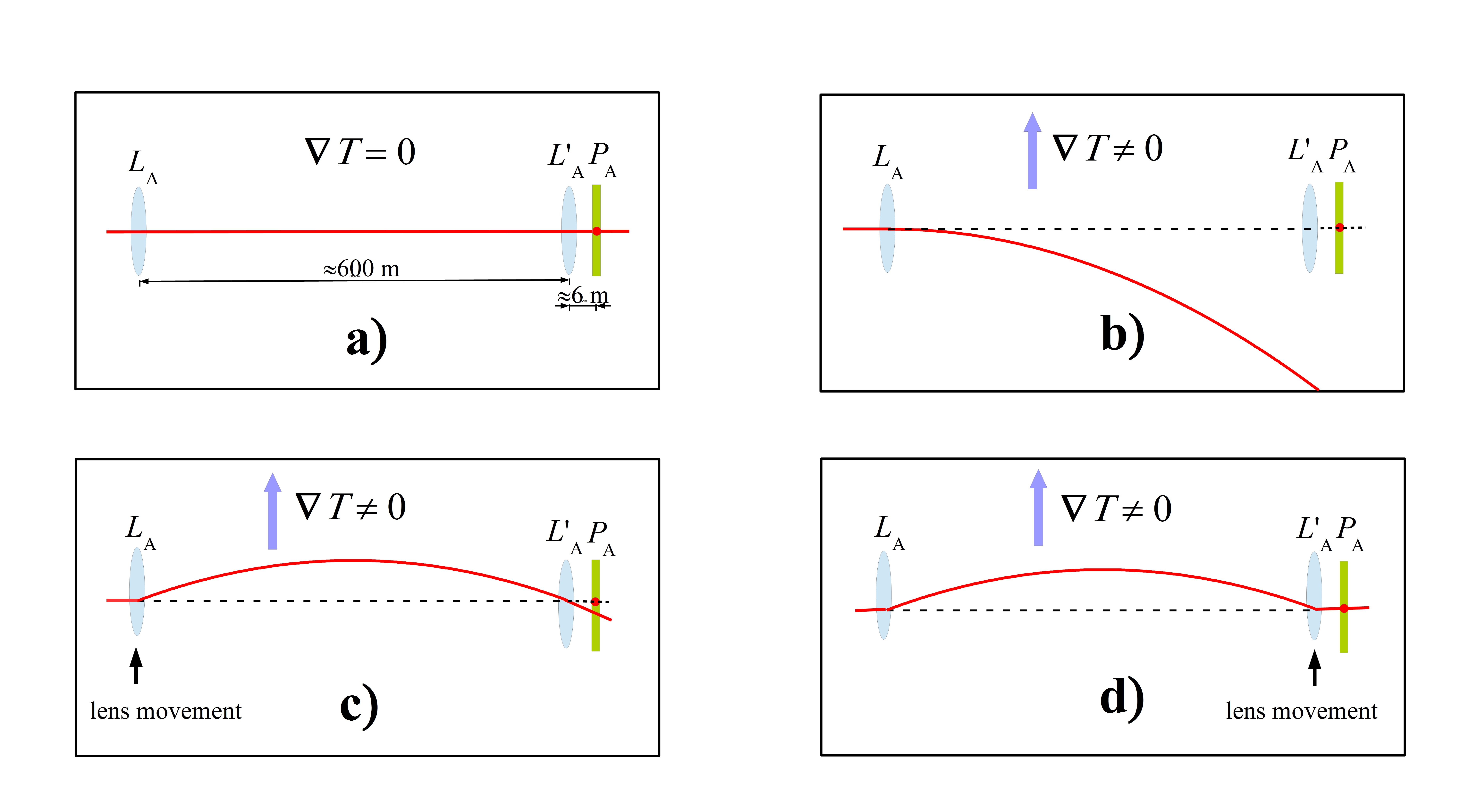}\caption{\label{figure:5-1} The average trajectory of beam \emph{I} in an
arm (the Alice arm) of the \emph{EGO} gallery with and without vertical
temperature gradient. The figure is not to scale but two characteristic
dimensions are given in figure (a). (a): The vertical temperature
gradient is not present: the average beam trajectory is a straight
horizontal line and the beam impinges at the centre of polarizer $P_{A}$.
(b) A vertical temperature gradient is present:  the reference beam
\emph{I} exhibits a parabolic trajectory. (c) A suitable movement
of lens $L_{A}$ deviate the transmitted beam and leads to a new parabolic
trajectory that passes through the centre of lens $L'_{A}$. Note
that, now, the incidence angle of the beam on lens $L'_{A}$ is not
zero and the transmitted beam does not impinge at the centre of polarizer
$P_{A}$. (d) A suitable movement of lens $L'_{A}$ deviate the transmission
beam toward the centre of polarizer $P_{A}$. }
\end{figure}

An interferometric method is used to equalize the optical paths from
the source of the entangled photons to polarizers $P_{A}$ and $P_{B}$.
The method exploits two reference beams (beams \emph{I} in figure
\ref{fig:3}) of wavelength $\lambda=681\,\mathrm{nm}$ and coherence
length $L_{A}=28.1\,\mathrm{\mu m}$ produced by a super luminous
diode (\emph{SLED} in figure  \ref{fig:3}). Due to the occurrence
of vertical temperature gradients up to $3\,\mathrm{\text{\textdegree C/m}}$
in the \emph{EGO} gallery produced by sunlight, it has been needed
to use two couples of different reference beams (beams\emph{ I} and
\emph{II}) in each arm of the interferometer. It can be easily shown
that an uniform vertical temperature gradient generates a vertical
gradient of the air refractive index that produces the same effect
as a diffused optical prism leading to a continuous deviation of the
optical beams \emph{I }and \emph{II} up to about $1\,\mathrm{m}$
at a $600\,\mathrm{m}$ distance.The full curve in Figure \ref{figure:5-1}(b)
shows the average trajectory of beam \emph{I }when a vertical temperature
gradient occurs. A parabolic shape of the trajectory is predicted
if the vertical temperature gradient is everywhere constant in the
gallery. Furthermore, the small non uniformity of the vertical gradient
of the air refractive index simulates a diffused cylindrical lens
that leads to astigmatism of the images. 

\begin{figure}
\centering{}\includegraphics[scale=0.7]{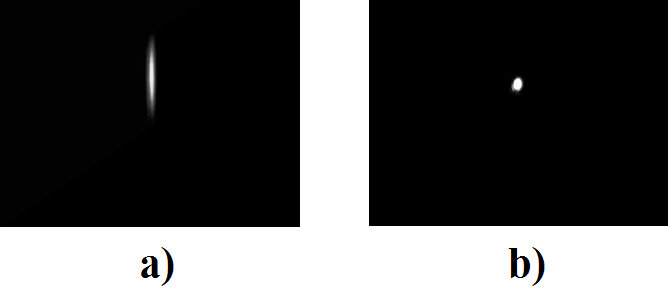}\caption{\label{figure:5} (a): The beam spot of the reference beam\emph{ I
}on polarizer $P_{A}$ in conditions of moderate sunlight and with
feedback \emph{OFF.} (b): The same spot with feedback \emph{ON}. The
diameter of the circular spot in \ref{figure:5}b) is 0.3 mm.}
\end{figure}

The accurate compensation of these effects is needed to collect a
great number of entangled photons on the photon counting detectors.
Beam \emph{I} follows the same optical path of the entangled photons
and provides an interferometric signal whilst beam \emph{II} is horizontally
displaced with respect to beam \emph{I} and allows us to compensate
the vertical deflections of the beams (up to $1\,\mathrm{m}$ at a
$600\,\mathrm{m}$ distance) produced by the air refractive index
gradients. The method to generate the reference beams has been greatly
improved with respect to that proposed in~\cite{Cocciaro_DICE2013}.
Here we obtain the reference beams \emph{I} and \emph{II} using the
beam displacer (Thorlabs BDY12U) in figure  \ref{fig:3} to split
the incident \emph{SLED} beam into two parallel beams at a $1.2\,\mathrm{mm}$
horizontal distance. The two beams pass through a beam splitter and
are focused on a transmission phase grating that produce +1 and -1
order diffracted beams with 35\% intensity with respect to the incident
beam and at the average diffraction angles $+2.43\text{\textdegree}$
and $-2.43\text{\textdegree}$ that are virtually coincident with
the maximum emission angles of the entangled photons ($\pm2.42\text{\textdegree}$).
The beam waists of the two reference beams spots on the optical grating
have a $0.3\thinspace\text{mm}$ diameter and behave as two sources
localized on the grating at a $1.2\thinspace\mathrm{mm}$ horizontal
distance. The optical rays emitted by these sources at the angles
$+2.43\text{\textdegree}$ and $-2.43\text{\textdegree}$ pass through
an achromatic lens having a $150\thinspace\mathrm{mm}$ focal length,
are reflected by a $565\thinspace\mathrm{nm}$ short pass dichroic
mirror (Chroma T565spxe) and produce 1:1 images of the grating spots
on the \emph{BBO} plates. Using a suitable optical method, the image
of the reference source \emph{I} on the \emph{BBO} plates is centred
with respect to the spot of the pump beam where the entangled photons
are generated. The procedure above ensures that the reference beams
\emph{I }outgoing from the \emph{BBO} plates are initially in phase
and are superimposed to the entangled photons. This provides the easy
alignment of the optical apparatus and the control of the optical
paths of the entangled photons. Achromatic lenses $L_{A},\thinspace L_{B},\thinspace L'_{A}$
and $L'_{B}$ have been built to have the same $6000\thinspace\mathrm{mm}$
focal length (within $\pm10\thinspace\mathrm{mm}$) for the pump laser
and for the \emph{SLED}. 1:1 images of the reference source \emph{I}
and of the entangled photons source are produced by lenses $L_{A},\thinspace L_{B},\thinspace L'_{A}$
and $L'_{B}$ on polarizers $P_{A}$ and $P_{B}$ at a $600\thinspace\mathrm{m}$
distance from the source. Reference beams\emph{ II} are horizontally
deflected by lenses $L_{A}$ and $L_{B}$ and produce two spots at
a horizontal distance of $12\thinspace\mathrm{cm}$ from the centres
of lenses $L'_{A}$ and $L'_{B}$ on two diffusing screens horizontally
adjacent to the lenses (see figure  \ref{fig:3}). Two optical objectives
collect the diffused beams and produce images of the spots on two
webcams. All lenses $L_{A},\thinspace L_{B},\thinspace L'_{A}$ and
$L'_{B}$ can be moved horizontally and vertically using Sigma Koki
\emph{PC} controlled motors. A labview program measures the position
of the beams spots on the webcams and produces feedback signals that
move lenses $L_{A}$ and $L_{B}$ to maintain the spots positions
fixed. In such a way also the reference beams \emph{I }(and the entangled
photons) always remain fixed at the centre of lenses $L'_{A}$ and
$L'_{B}$ (see Figure \ref{figure:5-1}(c)). A $1\thinspace\mathrm{cm}$
movement of lenses $L_{A}$ and $L_{B}$ produces a $1\thinspace\mathrm{m}$
displacement of the beams spots at a $600\thinspace\mathrm{m}$ distance.
The feedback procedure leads to a complete control of the slow drifts
of the beams but it cannot eliminate the rapid changes of the beam
trajectories occurring within a few seconds time. This leads to a
residual fluctuation of the beam spots at the centres of lenses $L'_{A}$
and $L'_{B}$ that is lower than $\pm30\thinspace\mathrm{mm}$. These
residual displacements are appreciably smaller than the radius of
the lenses ($75\thinspace\mathrm{mm}$), then all the entangled photons
are collected by them. The beams impinge at the centres of lenses
$L'_{A}$ and $L'_{B}$ but the incidence angles change with time,
due to the vertical refractive index gradient. Then, the images of
the source of beams \emph{I }(and of the entangled photons) that occur
on polarizers $P_{A}$ and $P_{B}$ do not remain fixed at the centre
of the polarizers (see figure \ref{figure:5-1}c)). To stabilize these
images at the centre of the polarizers, the reference beams \emph{I}
transmitted by the polarizers pass through two systems of cylindrical
lenses $CO_{A}$ and $CO_{B}$, are reflected by the $760\thinspace\mathrm{nm}$
long pass dichroic mirrors $DM_{A}$ and $DM_{B}$ (Chroma T760lpxr)
and impinge on two optical position control systems. Each optical
position control system produces two 1:1 images of the beam spots
occurring on the polarizers: one image is collected by a position
sensing detector (Thorlabs PDP90A) and the other by a webcam. Labview
feedback programs read the output of the position sensing detectors
and move lenses $L'_{A}$ and $L'_{B}$ to maintain fixed the position
of the beam spots at the centre of the two polarizers (see Figure
\ref{figure:5-1}(d)). In this case, too, slow drifts are completely
removed but not the rapid displacements of the spots from the polarizers
centres. These latter residual fluctuations remain always restricted
below $\pm0.4\thinspace\mathrm{mm}$. Other labview feedback programs
acquire the images of the webcams and measure the astigmatism of the
images induced by non uniformities of the refractive index gradients.
Each system of cylindrical lenses $CO_{A}$ and $CO_{B}$ is composed
by a fixed cylindrical lens and a movable cylindrical lens that provide
an effective cylindrical lens with a variable focal length. Suitable
feedback signals generated by the labview program move the motorized
cylindrical lenses to correct the astigmatism of the images. These
procedures ensure that the spots of the reference beams \emph{I} remain
virtually fixed at the centre of polarizers $P_{A}$ and $P_{B}$
with a circular shape having a $\approx0.3\thinspace\mathrm{mm}$
diameter. Two images of the spot of the reference beam on polarizer
$P_{A}$ are shown in figures \ref{figure:5}a) and \ref{figure:5}b).
Figure \ref{figure:5}a) shows the spot for moderate sunlight with
feedback \emph{OFF}, whilst figure \ref{figure:5}b) shows the same
image with feedback \emph{ON}. It must be remarked that our method
stabilizes the spots of the 681 nm reference beams but the wavelengths
of the entangled photons ($790\thinspace\mathrm{nm}-830\thinspace\mathrm{nm}$)
are different from those of the sled beam. However, the differences
of the air refractive indices corresponding to the reference beams
and to the entangled photons are very small and it can be shown that
also the spots of the entangled photons on the polarizers always remain
very close to the centres of the polarizers within $0.4\thinspace\mathrm{mm}$.
In conclusion, our compensation procedure maintains the spot of the
entangled photons restricted to a circular region close to the centre
of the polarizers with a small diameter ($\approx0.6\thinspace\mathrm{mm}$)
and ensures that virtually all the entangled photons passing through
the polarizers are collected by the photon counting detectors also
in conditions of great sunlight.

\subsection{\label{reference beams-1} Equalization of the optical paths.}

\begin{figure}
\centering{}\includegraphics[scale=0.06]{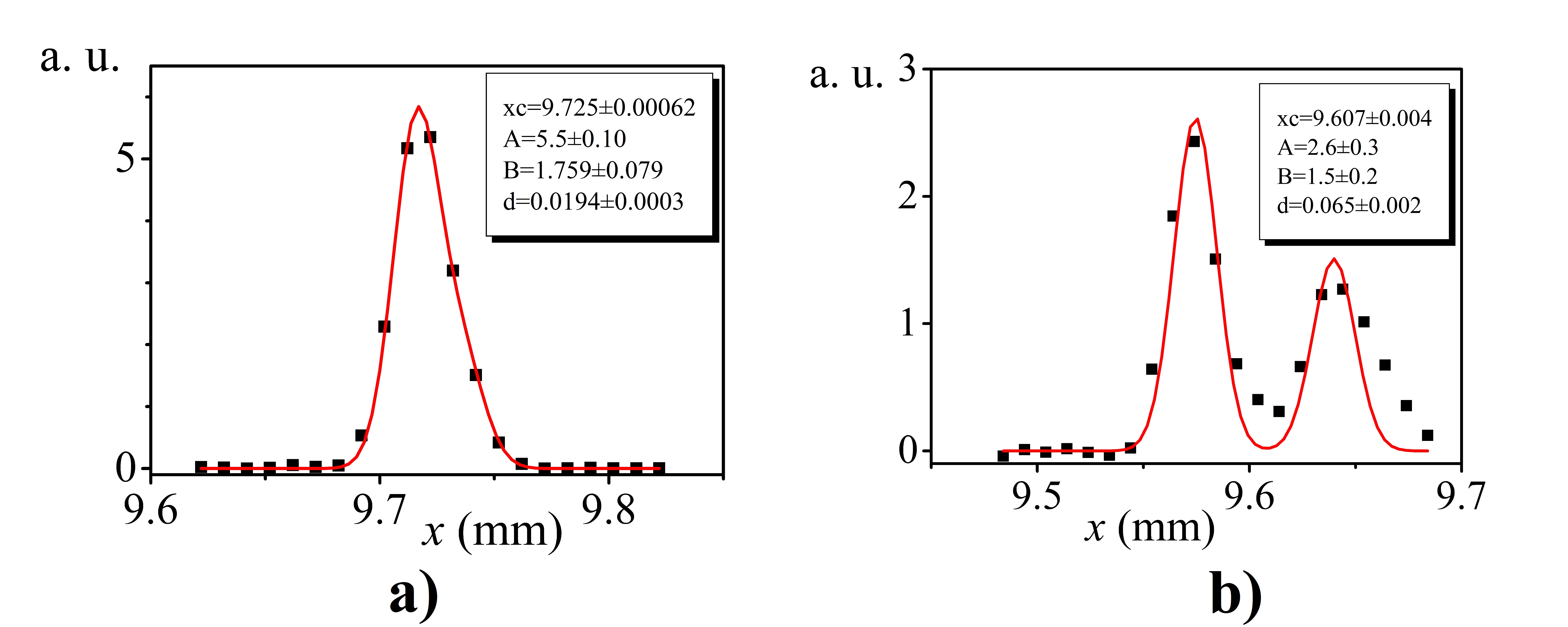}\caption{\label{fig:6} (a): The peak to peak of the output voltage of photodetector
\emph{Ph} during a Summer night versus the position \emph{x} of the
polarizer. Similar results are obtained in conditions of covered sky.
The full line is the best fit with two slightly shiften Gaussian profiles
$y=A\exp\left\{ \left[(x-x_{c}-d/2)/w\right]^{2}\right\} +B\exp\left\{ \left[(x-x_{c}+d/2)/w\right]^{2}\right\} $
where \emph{w} is fixed to the value $w=0.020\thinspace\mathrm{mm}$
that characterizes the \emph{SLED }source and \emph{A}, \emph{B},$x_{c}$
and \emph{d} are fitting parameters. $x_{c}$ and \emph{d} are measured
in millimeters whilst \emph{A} and \emph{B} are in arbitrary unities.
(b): The same dependence but in conditions of maximum sunligth. The
two fitting Gaussians show a much greater separation due to the much
greater amplitude of the optical paths fluctuations and the quality
of the fit is poor due to noise effects induced by the strong air
tourbolence. }
\end{figure}

To equalize the optical paths we exploit the reflections of the reference
beams \emph{ I} from polarizers $P_{A}$ and $P_{B}$. The reflected
beams come back on the same path forward and impinge at the angles
$+2.43\text{\textdegree}$ and $-2.43\text{\textdegree}$ on the optical
phase grating where diffraction occurs again. The output beams that
are diffracted orthogonally to the grating are reflected by the beam
splitter and impinge on photodetector \emph{Ph} where interference
occurs. Air density fluctuations inside the \emph{EGO} gallery induce
oscillations of the optical path difference and, thus, an oscillating
output voltage of the photodetector. The variations of the optical
path differences are always greater than the optical wavelength and
the output voltage oscillates from a minimum value (destructive interference)
toward a maximum value (constructive interference). The peak to peak
amplitude $V_{pp}$ is measured by a simple electronic circuit. $V_{pp}$
is maximized when the path difference is zero whilst $V_{pp}$ tends
to vanish if the path difference becomes greater than the \emph{SLED}
coherence length $L_{c}=28.1\,\mathrm{\mu m}$. Polarizer $P_{B}$
is moved by a precision linear motorized stage (Physik Instruments
M-406.22s) that is controlled by a \emph{PC} through a labview program
that generates a sweep of the $P_{B}$ position and acquires the corresponding
$V_{pp}$ values. The typical dependence of $V_{pp}$ on the polarizer
position \emph{x} during a summer night is shown in figure  \ref{fig:6}
(a) whilst the dependence during a summer day at the maximum sunlight
is shown in figure  \ref{fig:6}(b). Note that the curve in figure
 \ref{fig:6}(b) has a two bells profile. This behaviour can be explained
assuming that the path difference oscillates with time around the
average value $\Delta d_{av}$ with a mean oscillation amplitude \emph{A}.
In these conditions, the typical Gaussian behaviour due to the finite
coherence length of the \emph{SLED} is expected to split into two
nearly Gaussian profiles at distance 2\emph{A}. The central point
between the two Gaussian peaks corresponds to the position of polarizer
$P_{B}$ where the average optical path difference is zero whilst
the semi-distance between the two Gaussian Maxima corresponds to the
average amplitude \emph{A} of the fluctuations of the paths difference.
The Full lines in figures  \ref{fig:6}(a) and \ref{fig:6}(b) are
the labview best fits of the experimental results with two Gaussians
having the width $w=0.020\thinspace\mathrm{mm}$ that characterizes
the \emph{SLED} source. From these best fits we deduce that the main
amplitude of the oscillations of the path difference is smaller than
$10\thinspace\mu m$ during night but it becomes $33\thinspace\mu m$
at the maximum sunlight.

\begin{SCfigure}[50]
 \centering
 \includegraphics[width=0.5\textwidth]{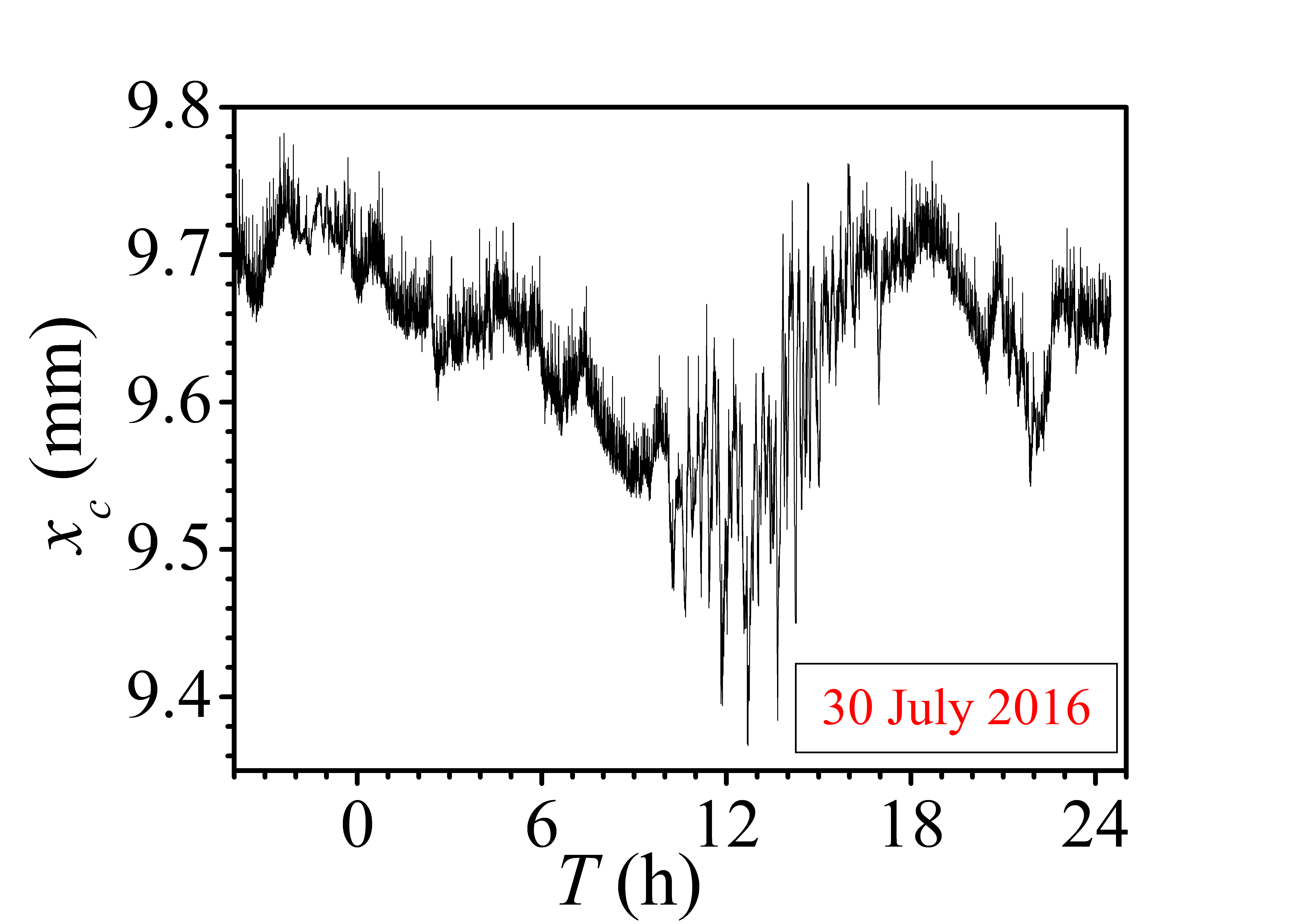}
 \hspace{0.05 in}
 \caption{\protect\rule{0ex}{12ex}Variation of the equalization position $x_{c}$ of the  motorized polarizer during an entire Summer day. The maximum variation during the entire day is about 0.4 mm. Note the sharp and high amplitude variations due to the air turbulence during the maximum sunlight hours (from 10 h to 16 h). }
\label{fig:7}
\end{SCfigure}

The labview feedback program operates in this way: first of all a
large amplitude sweep is made to localize the central point $x_{c}$
between the two Gaussian, then the sweep amplitude is reduced to $200\thinspace\mu m$
around $x_{c}$ and the new $x_{c}$ value is memorized and plotted.
This latter procedure with a $200\thinspace\mu m$ sweep is repeated
continuously each $15\,\mathrm{s}$ for the entire measurement time
(24 hours) and, thus, the difference between the optical paths remains
restricted to $\Delta d=\pm100\thinspace\mathrm{\mu m}$ at each time.
The time-dependence of $x_{c}$ due to the temperature variations
for an entire Summer day is shown in figure \ref{fig:7}.

Note that the complete interference pattern is somewhat more complex
than the small portion shown in figures  \ref{fig:6} since the \emph{LPNIR}
Thorlabs polarizers are made by a thin polarizing film ($280\thinspace\mathrm{\mu m}$
thickness) sandwiched between two glass plates. We have measured with
accuracy the thickness of the glasses and we have found that their
values are $904\thinspace\mathrm{\mu m}\pm5\thinspace\mathrm{\mu m}$.
Due to the sandwich shape of the polarizers, there are many interfaces
giving reflected beams that can interfere. Also in the optimal night
conditions we observe five main interference peaks that are separated
each from the other by a distance about $1.5\thinspace\mathrm{m}\mathrm{m}$.
After a careful analysis, the central peak has been clearly identified
as that which corresponds to the equalization of optical paths from
the source to the polarizing thin layers. Then, our analysis has been
restricted only to the central peak region corresponding to figures
 \ref{fig:6}.

As shown in Section \ref{sec:Introduction}, one of the most important
parameters of our experiment is the uncertainty $\Delta d$ on the
equalization of the optical paths. Here we resume the main contributions
to this uncertainty (see also~\cite{Cocciaro_DICE2013}) :

a) the above described uncertainty due to the motor sweep $\Delta d_{sweep}=100\thinspace\mathrm{\mu m}$
that corresponds to a half of the total sweep excursion, 

b) the uncertainty $\Delta d_{pol}$ due to the finite thickness ($280\thinspace\mu m$)
of the \emph{LPNIR} Thorlabs polarizers layers. Since the extinction
ratio of these polarizers at the entangled photons wavelengths is
greater than $10^{5}$, then we can estimate that 99\% of photons
with orthogonal polarization are adsorbed in a layer having a thickness
$\approx120\thinspace\mathrm{\mu m}$ and we can assume the corresponding
uncertainty value $\Delta d_{pol}=120\thinspace\mathrm{\mu m}$.

c) The optical paths are equalized using the reference beams \emph{I
}that have not the same wavelength of the entangled photons. If the
temperature would be uniform in the \emph{EGO} gallery and the thickness
of lenses $L_{A},\thinspace L_{B},\thinspace L'_{A}$ and $L'_{B}$
would be the same, also the entangled photons paths would be automatically
equalized. This is no more true if there is a average temperature
difference $\Delta T$ between the two arms of the interferometer
or if there are differences between the thicknesses of the lenses.
Calculations of these effects are somewhat complicate and need the
knowledge of the temperature dependence of the refractive indices
of the air and of the lenses at different wavelengths and the knowledge
of the thickness differences between lenses $L_{A},\thinspace L_{B},\thinspace L'_{A}$
and $L'_{B}$. The lenses thickness differences have been measured
to be smaller than 0.1 mm and they do not affect appreciably the uncertainty.
It results from the calculations that an average temperature difference
$\Delta T=1\text{\textdegree C}$ between the two arms of the interferometer
produces an optical paths difference slightly smaller than $10\thinspace\mathrm{\mu m}$.
Since the horizontal temperature differences in the gallery are always
smaller than 2-3 degrees, we get $\Delta d_{\Delta T}<30\thinspace\mathrm{\mu m}$. 

d) In our experiment we detect entangled photons with wavelengths
from $790\thinspace\mathrm{nm}$ toward $830\thinspace\mathrm{n}\mathrm{m}$.
Due to the optical dispersion, photons of different wavelengths see
different optical paths in air and in the lenses, although this latter
contribution is negligible. The difference of the optical paths due
to the air optical dispertion  is given by $\Delta d_{disp}=\frac{\partial n}{\partial\lambda}\Delta\lambda d$,
where \emph{d} is the distance from the source to polarizers (600
m) and $\Delta\lambda$ = 40 nm is the bandwidth of the bandpass filters
and \emph{n} is the air refractive index. Substituting in the expression
of $\Delta d_{disp}$ the value $\frac{\partial n}{\partial\lambda}=5.87\times10^{-9}nm^{-1}$
calculated using the Ciddor equation~\cite{CiddorEquation} at room
conditions and with humidity = 50\% and $CO_{2}$ = 450 micromol/mol,
we get $\Delta d_{disp}=$ 144 $\mu m$. The resulting uncertainty
in the optical paths differences is, then

\begin{equation}
\Delta d=\sqrt{\Delta d_{sweep}^{_{^{2}}}+\Delta d_{pol}^{_{^{2}}}+\Delta d_{\Delta T}^{_{^{2}}}+\Delta d_{disp}^{_{^{2}}}}=215\thinspace\mu m.\label{eq:betamin-1}
\end{equation}

\section{\label{sec:Critical-points-and} Preliminary experimental results.}

In this Section we report some very preliminary experimental results
concerning the \emph{EPR} measurements. The main objective of these
measurements is to verify that the experimental method provides accurate
measurements of the\emph{ EPR} correlations with very small acquisition
times of the coincidences. As shown in the Introduction, the presence
of superluminal communications can be detected looking at the time
dependence of the two correlation parameters $S_{max}$ and $S_{min}$.
Counts $N_{A}$ and $N_{B}$ of photons transmitted by polarizers
$P_{A}$ and $P_{B}$ are detected and back-ground counts due to unwanted
external light and to dark noise of the detectors are subtracted.
Furthermore, the statistic spurious coincidences $N_{sp}=N_{A}N_{B}\delta't$
are subtracted from the measured coincidences, where $\delta't$ denotes
the pulse duration $\delta't=25\thinspace\mathrm{ns}$ of the output
pulses generated by the photon counting modules. Numbers $N(\alpha_{A},\alpha_{B})$
of the measured coincidences that appear in the expressions of the
correlation parameters $S_{max}$ and $S_{min}$ are affected by the
time-variations of the transmission coefficients in the two arms.
The main causes of a variation of the transmission coefficients are:
the occurrence of residual displacements of the transmitted beams
that are not completely eliminated by the feedback method that reduce
the collection efficiency of the entangled photons; the variation
of the air relative humidity that induces a change of the light adsorption.
In fact, counts $N_{A}(\alpha_{A})$ and $N_{B}(\alpha_{B})$ and
coincidences $N(\alpha_{A},\alpha_{B})$ are related to the transmission
coefficients according to relations (\ref{eq:Smax-1-1}):
\begin{equation}
\begin{array}{l}
N_{A}(\alpha_{A})=N\tau_{A}(\alpha_{A})\epsilon_{A}(\alpha_{A})p_{A}(\alpha_{A})\\
N_{B}(\alpha_{A})=N\tau_{B}(\alpha_{B})\epsilon_{B}(\alpha_{B})p_{B}(\alpha_{B})\\
N(\alpha_{A},\alpha_{B})=N\tau_{A}(\alpha_{A})\epsilon_{A}(\alpha_{A})\tau_{B}(\alpha_{B})\epsilon_{B}(\alpha_{B})p(\alpha_{A},\alpha_{B})
\end{array}\label{eq:Smax-1-1}
\end{equation}
where \emph{N} is the number of generated entangled photons, $\tau_{A}(\alpha_{A})$
and $\tau_{B}(\alpha_{B})$ are the transmission coefficients, $\varepsilon_{A}(\alpha_{A})$
and $\varepsilon_{B}(\alpha_{B})$ are the efficiencies of photon
counting detectors, $p_{A}(\alpha_{A})$ and $p_{B}(\alpha_{B})$
are the probabilities that photons \emph{a }and \emph{b} pass through
polarizers $P_{A}$ and $P_{B}$ (they are $p_{A}(\alpha_{A})=p_{B}(\alpha_{B})=1/2$
for the entangled state) and $p(\alpha_{A},\alpha_{B})$ is the joint
probability. We see that dividing the coincidences counts $N(\alpha_{A},\alpha_{B})$
for the product $N_{A}(\alpha_{A})N_{B}(\alpha_{B})$ and multiplying
for the product of the average values of $\bar{N}_{A}(\alpha_{A})$
and $\bar{N}_{A}(\alpha_{A})$ one obtains a coincidences number that
is no more affected by changes of the transmission coefficients and
of the photodetectors efficiencies. Here below we will indicate by
$N(\alpha_{A},\alpha_{B})$ the coincidences corrected according to
the procedure outlined above. The correlations parameters $S_{max}$
and $S_{min}$ are obtained repeating measurements of coincidences
$N(\alpha_{A},\alpha_{B})$ with the proper values of angles $\alpha_{A}$
and $\alpha_{B}$ that appear in equations (\ref{eq:Smax}) and (\ref{eq:Smin}).
According to equations (\ref{eq:Smax}), (\ref{eq:Smin}) and (\ref{eq:NTOT}),
12 different couples of values $\alpha_{A}$ and $\alpha_{B}$ have
to be selected rotating the motorized polarizers $P_{A}$ and $P_{B}$.
The rotations of polarizers $P_{A}$ and $P_{B}$ are controlled by
a \emph{PC} through a labview program that operates in this way: a
couple of angles $\alpha_{A}$ and $\alpha_{B}$ is set (for instance
the first angles 45\textdegree{} and 67.5\textdegree{} of the first
contribution in equation (\ref{eq:Smax})), then the polarizers axes
are rotated until they reach the setted angles. The corresponding
numbers of coincidences $N(\alpha_{A},\alpha_{B})$ in the acquisition
time $\delta t$ are measured. Then, angles are changed according
to equations (\ref{eq:Smax}), (\ref{eq:Smin}) and (\ref{eq:NTOT})
and the corresponding values of the coincidences are measured. When
all the 12 values of coincidences that are needed to calculate parameters
$S_{max}$ and $S_{min}$ have been measured, the labview program
calculates these correlation parameters. Unfortunately, the average
time that is needed to rotate the polarizers is of the order of 8
seconds and, thus, the duration of a single measurement of $S_{max}$
and $S_{min}$ requires a time $\Delta t\approx100\thinspace\mathrm{s}$
that is much larger than the acquisition time $\delta t$ of coincidences.
Then, the maximum superluminal velocity $\beta_{t,min}$ that can
be detected in the present experiment is not limited by the acquisition
time $\delta t$ of the coincidences but by the much larger effective
acquisition time $\Delta t=100$ s.
\begin{figure}
\centering{}\includegraphics[scale=0.06]{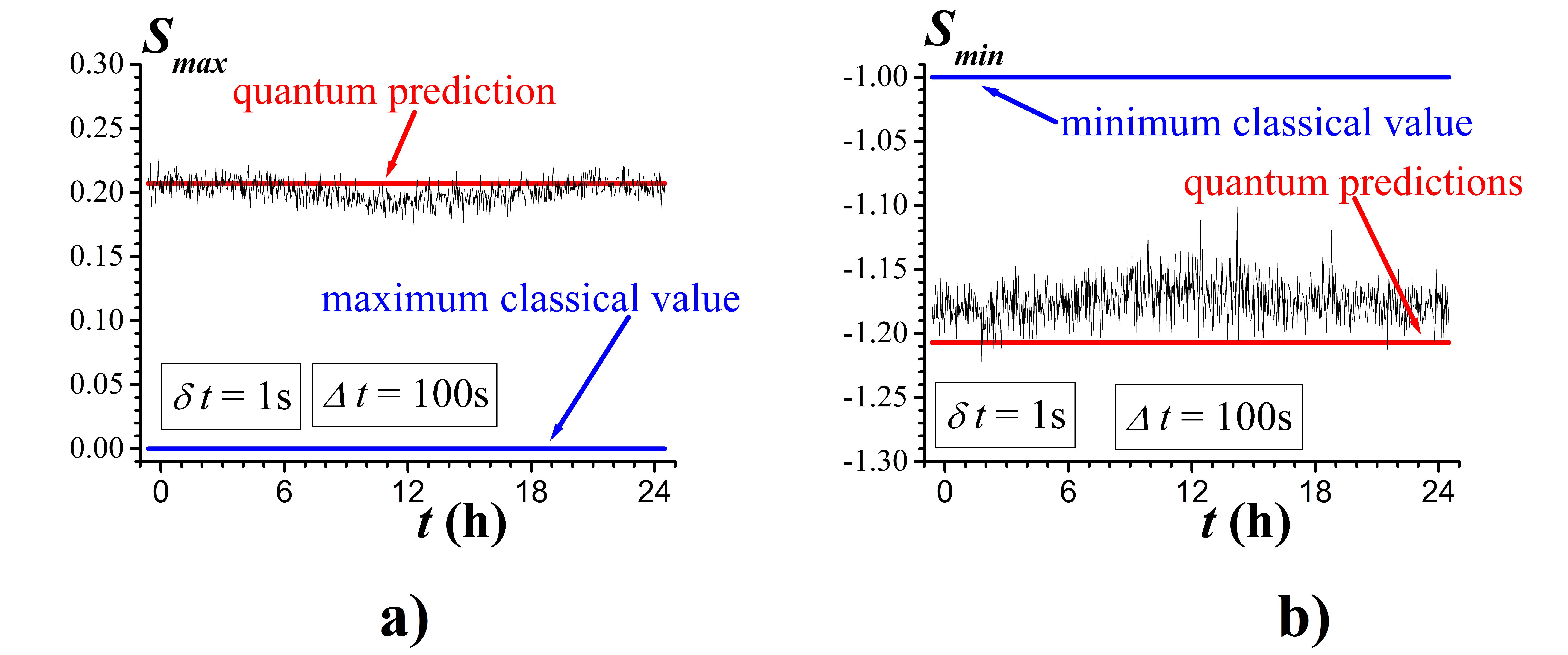}\caption{\label{fig:8} (a): Correlation parameter $S_{max}$ versus the sidereal
time. The full red line indicates the dependence predicted by Quantum
Mechanics, whilst the blue lower line represents the maximum value
allowed by local models of \emph{QM}. (b): The same dependence but
for the other parameter $S_{min}$. Note that parameter $S_{min}$
is much more noisy because it is more sensitive to fluctuations of
the coincidences numbers. Note the presence of a systematic effect
of the sunlight in the interval between 9 h and 18 h.}
\end{figure}
 Figures  \ref{fig:8}a) and \ref{fig:8}b) show the values of the
correlation parameters $S_{max}$ and $S_{min}$ versus time during
a sidereal day when the coincidences acquisition time was $\delta t=$1
s but the effective acquisition time was $\Delta t=100\thinspace\mathrm{s}$.
The full horizontal lines in figures  \ref{fig:8}a) e and \ref{fig:8}b)
correspond to the prediction of \emph{QM} and to the maximum (figure
 \ref{fig:8}a)) and minimum (figure  \ref{fig:8}b)) values allowed
by local theories. According to the Introduction, parameter $S_{max}$
would become lower than 0 and $S_{min}$ would become greater than
- 1 at two times each sidereal day if the superluminal signals have
velocities lower than $\beta_{t,min}$. This behaviour is not observable
in figures  \ref{fig:8}a) and \ref{fig:8}b) and, thus, we can conclude
that, if superluminal signals are responsible for \emph{QM} correlations,
then the superluminal velocities are greater than the maximum measurable
values ($\beta_{t}$$>$$\beta_{t,min}$). The results in figures
\ref{fig:8}a) and \ref{fig:8}b) were obtained using the coincidences
acquisition time $\delta t=1\thinspace\mathrm{s}$ but we have verified
that sufficiently accurate results are also obtained using the much
smaller acquisition time $\delta t=0.1\thinspace\mathrm{s}$ where
the relative statistical noise increases by a factor $\sqrt{10}$.
Notice that parameter $S_{min}$ exhibits much greater fluctuations
than $S_{max}$ and, thus, this latter parameter provides a much more
accurate test of the \emph{EPR} correlations. This behaviour is probably
due to the fact that the absolute value of $S_{max}$ is about six
times lower than that of $S_{min}$. For this reason the planned final
measurements will be made using the $S_{max}$ parameter alone that
is affected by a much smaller noise. Furthermore, it is important
to remark that the measurements shown in figures  \ref{fig:8}a) and
\ref{fig:8}b) were obtained in a July day (30 July 2016) with very
strong sunlight. The residual noisy effects due to sunlight are evident
looking at the experimental points between times \emph{t }= 9 h and
\emph{t }= 18 h in the figures. All these effects are absent in conditions
of fully covered sky.

\begin{figure}
\centering{}\includegraphics[scale=0.06]{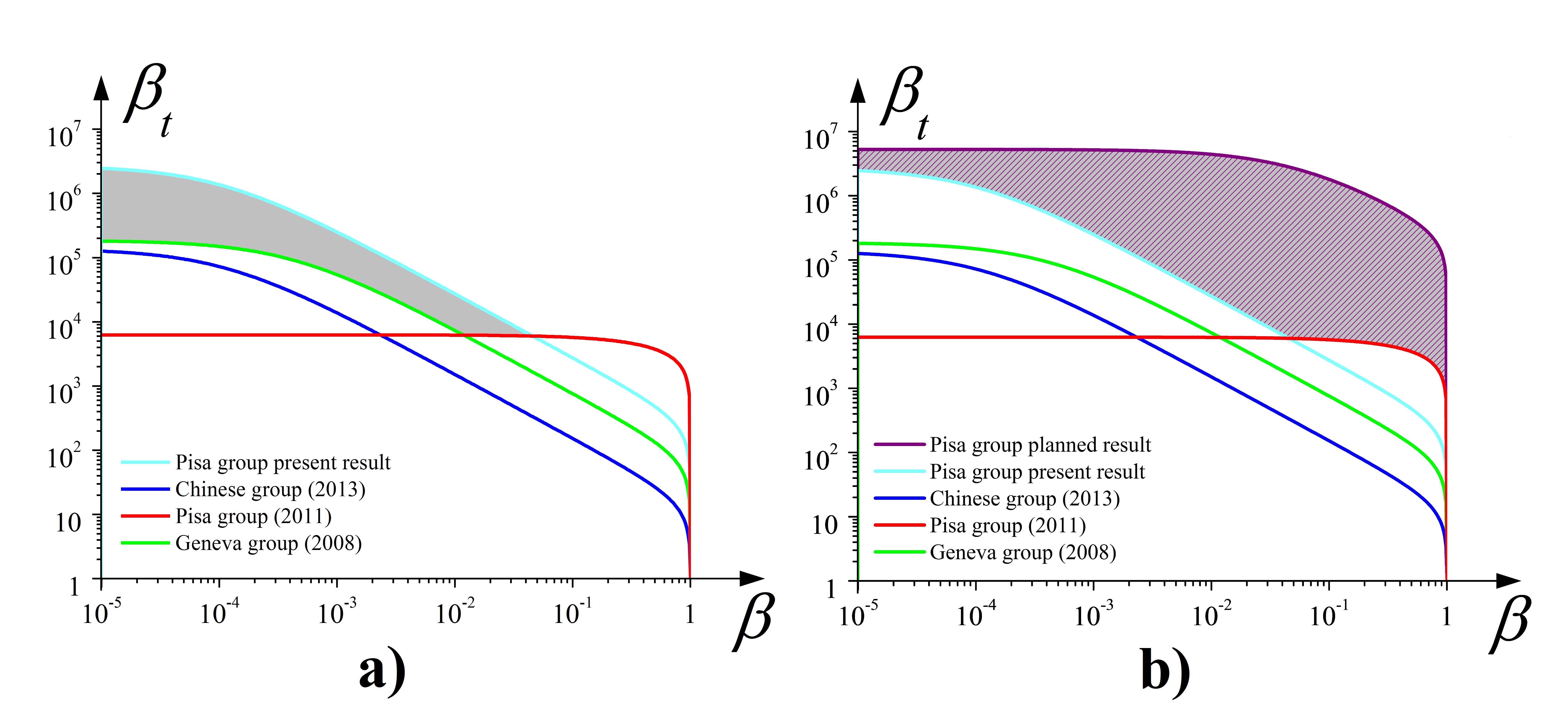}\caption{\label{fig:9} (a): The reduced superluminal velocity $\beta_{t}$
versus the reduced velocity $\beta$ of the preferred frame. The curves
in the figure represent the lower bounds $\beta_{t,min}$ that have
been established in the previous experiments and in the present one.
The filled region represents the new region of superluminal velocities
that are investigated here. (b) The filled region represents the new
region of superluminal velocities that should be explored in the final
experiment with the acquisition time $\delta t$ = 0.1 s. }
\end{figure}
Substituting the effective acquisition time $\Delta t=100\thinspace\mathrm{s}$
in place of $\delta t$ in equation (\ref{eq:betamin}) with the uncertainty
$\Delta d=215\thinspace\mathrm{\mu m}$, we obtain the lower bound
$\beta_{t,min}$ that corresponds to our preliminary results. In figure
\ref{fig:9} a) we show the lower bounds already found in some previous
experiments~\cite{Salart_nature_2008,Cocciaro_PLA_2011,Cinesi_PhysRevLett2013}
together with that obtained here. The filled region represents the
new region of superluminal velocities investigated here. In figure
 \ref{fig:9} b) we show also the planned values of $\beta_{t,min}$
that should be obtained in our final experiment with a $0.1\thinspace\mathrm{s}$
effective acquisition time. The filled region in figure  \ref{fig:9}
b) corresponds to the new region of superluminal velocities that will
become accessible in the final experiment. The experimental method
that will be used to bypass the problems related to the polarizers
movement will be briefly outlined in the Conclusions below.

\section{\label{sec:Conclusions}Conclusions}

In the present paper we have developed an accurate and stable method
to equalize the optical paths of the entangled photons over a kilometric
distance. Due to vertical gradients of the air refractive index in
the \emph{EGO} gallery induced by sunlight it has been needed to greatly
modify the experimental apparatus proposed in~\cite{Cocciaro_DICE2013}
introducing a complex feedback procedure to correct the deviations
of the beams and the astigmatism of the images. In such a way we were
able to obtain two virtually stable 1:1 images of the source of entangled
photons at the centre of two polarizers lying at a distance of $\approx600\thinspace\mathrm{m}$
from the source. This ensures that virtually all the entangled photons
transmitted by the polarizers are collected by the photon counting
detectors also in the unfavourable conditions of maximum sunlight.
The interference method used to equalize the optical paths exploits
two reference beams reflected by the polarizers. The reference beams
follow the average paths of the entangled photons. The method to produce
the reference beam has been greatly improved with respect to the original
project~\cite{Cocciaro_DICE2013} thanks to the use of a suitable
optical grating. The new method automatically ensures that the reference
beams are superimposed to the entangled ones and that they have the
same phase at the \emph{BBO} plates without using the complex equalization
procedure outlined in our original paper~\cite{Cocciaro_DICE2013}.
Finally, a suitable design of the optical components and the use of
the compensation procedure developed by the Kwiat group~\cite{Kwiat_OptExpr_2005,Kwiat_OptExpr_2007,Kwiat_OptExpr_2009}
provides a great number of measured coincidences and makes possible
to use a very short acquisition time $\delta t=0.1\thinspace\mathrm{s}$.
Using this experimental apparatus we have continuously measured the
correlation parameters $S_{max}$ and $S_{min}$ for an entire sidereal
day to obtain some very preliminary results. Our experimental results
are greatly affected by the long time that is needed to rotate the
polarizers that leads to an effective acquisition time $\Delta t=100\thinspace\mathrm{s}$
much greater than the minimum acquisition time of coincidences $\delta t=0.1\thinspace\mathrm{s}$.
For this reason, the new explored region of the velocities of the
superluminal signals investigated here (see figure  \ref{fig:9} a))
is much smaller than the planned one (see figure  \ref{fig:9} b)).
Due to the strong vertical temperature gradients it has not been possible
to perform the experiment along the true East-West direction that
was proposed in our previous paper. In the present experiment the
measures have been performed in the so called ``East-West'' gallery
of \emph{EGO} that makes the angle $\gamma=18\text{\textdegree}$
with the actual East-West axis. Then, only a 95\% portion of the celestial
sphere is accessible with our experimental apparatus.

In order to become insensitive to the long time needed to rotate the
polarizers and to reach an effective acquisition time $\delta t=0.1\thinspace\mathrm{s}$,
it is needed to fully change the acquisition method. With this improved
method we should obtain the planned results in figure  \ref{fig:9}
b). Here we describe only the main idea of the new method. Using a
NTP+PTP GPS Network Time Server (TM2000A) and knowing the values of
the difference \emph{UTC}-\emph{UT1} provided in the Web by \emph{IERS}~\cite{IERS}
we will be able to synchronize the measurements of the coincidences
with the Earth rotation angle with respect to the fixed stars within
a few milliseconds uncertainty. That accurate synchronization of the
measurements cannot be obtained using a \emph{PC} but requires the
use of a real time acquisition. This will be obtained using a \emph{Compact
DAQ }(National Instruments 9132) in place of the \emph{PC} to acquire
the coincidences and a Real Time Labview program to control any aspect
of the acquisition. In this way, will be possible to acquire the coincidences
in successive days at the same Earth rotation times. The measurement
method will follow the steps below: i) the real time labview program
rotates polarizers to reach the angles $\alpha_{A}$ and $\alpha_{B}$
that correspond to the first contribution $N(45\text{\textdegree},\thinspace67.5\text{\textdegree})$
in the expression of $S_{max}$ in equation (\ref{eq:Smax}). ii)
When the Earth rotation angle reaches a well defined value, the acquisition
of coincidences starts and $2^{20}$ values of coincidences are acquired
in a full Earth rotation day. The successive day, the real time labview
program sets the polarizers angles to the values corresponding to
the second term in the expression of $S_{max}$ and the acquisition
of coincidences will start with sidereal synchronism with the measurements
of the previous day. The same procedure will be repeated until all
the contributions that are present in the expression of $S_{max}$
have been obtained. With this procedure, the effective acquisition
time coincides with the coincidences acquisition time $\delta t$
and the planned region of superluminal velocities (filled region in
figure \ref{fig:9} b)) should become accessible. If the revolution
 motion of the Earth around the Sun and the precession and nutation
of the Earth axis would be absent, the losses of quantum correlations
should occur exactly at the same earth rotation angles each sidereal
day and, thus, one could utilize the coincidences $N(\alpha_{A},\alpha_{B})$
measured in successive days at the same earth rotation angles to calculate
the time dependence of the correlation parameter $S_{max}$ using
equations (\ref{eq:Smax}) and (\ref{eq:NTOT}). Unfortunately, the
analysis of the experimental data will be much more complex due to
the revolution motion of the earth around the Sun and to the precession
and nutation motions. In fact, the losses of quantum correlations
should occur when the relative velocity $\vec{V}$ of the preferred
frame with respect to the Earth frame becomes orthogonal to the \emph{A}-\emph{B}
axis. Due to the revolution motion of the earth around the Sun and
to the precession and nutation motions, the angle between the relative
velocity $\vec{V}$ of the preferred frame and the \emph{A}-\emph{B}
axis is not a true periodic function having the period of the Earth
rotation and, thus, the orthogonality condition will be not exactly
satisfied at the same Earth rotation angles in different days. Then,
the analysis of the experimental results will need more complex procedures
that will be not discussed here.

\section*{Acknowledgements}

We acknowledge Marco Bianucci for the realization of many electronic
devices and for a great number of helpful and fundamental suggestions.
We also thank the European Gravitational Observatory of Cascina (Italy)
that host our experiment and, in particular, the director Federico
Ferrini and Franco Carbognani and Stefano Cortese for their valuable
support and for their  kindness. We also thank the Istituto di tecnologie
della comunicazione, dell'informazione e della percezione (S. Anna)
for giving us two LCM1555EW4932-64 of Nortel Networks. Finally we
thank Nicolas Gisin for the useful discussions on the topics presented
above. This work was supported by La Fondazione Pisa. 

\bibliographystyle{iopart-num}
\bibliography{mybibmini}

\end{document}